\documentstyle[twocolumn,pra,aps,psfig]{revtex}

\begin{document}

\draft

% for two column  activate the line below...
\twocolumn[\hsize\textwidth\columnwidth\hsize\csname @twocolumnfalse\endcsname

\title{Li$_5$ as a pseudorotating planar cluster}

\author{R. Kawai and J. F. Tombrello}
\address{Department of Physics, University of Alabama at Birmingham,
Birmingham, Alabama 35294}

\author{J. H. Weare}
\address{Department of Chemistry, University of California, San Diego,
La Jolla, California 92093}

\date{\today}

\maketitle

\begin{abstract}
A pseudorotating state of Li$_5$ observed in a recent EPR experiment
is investigated using the local density functional method.  The calculated
isotropic spin population indicates that a planar $C_{2v}$ structure
is more consistent with the experimental result than a suggested trigonal
bipyramid.
A new pseudorotating state based on the planar structure is proposed.
\end{abstract}


% end of single column
]

\narrowtext

Despite a decade of theoretical investigation,
the geometric structures of small metallic clusters
are poorly understood. This is partly
due to the lack of a
reliable experimental technique and partly to the
inadequate accuracy of {\it ab initio}
calculations.  Only a few geometries have been
experimentally determined.  Spectroscopic data for the systems larger
than the trimer are
too complicated to extract structural information.
On the other hand, highly accurate configuration-interaction (CI)
calculations
are too expensive even for small clusters when
geometric optimization is required.
A solution to this challenging problem is to calculate spectroscopic data
or other observables using an accurate {\it ab initio} method for candidate
geometries and to eliminate structures that are inconsistent with
the experimental data.  Currently available {\it ab initio} methods
are accurate enough to distinguish quantitative differences in spectra due
to different structures\cite{KFK91}.
Recently, geometries of a few small lithium clusters
have been partially determined in this way\cite{BCDW90,DBKB91,BKBC92}.

For Li$_3$, both matrix-isolated\cite{GaLi83,HSM84}
and cluster-beam\cite{Hish92} electron paramagnetic resonance (EPR) spectra
showed that
all three nuclei are
magnetically equivalent, indicating that Li$_3$ is an equilateral
triangle ($D_{3h}$).  On the other hand, theoretical calculations
predicted two nearly degenerate energy minima,$^2A_1$ and $^2B_2$ states,
corresponding to $C_{2v}$ isosceles triangles.  This disagreement was
explained by pseudorotation around the $D_{3h}$ symmetry
point\cite{GeSc78,MCB83}.
Recently, extremum structures of the $^2A_1$ and $^2B_2$ states were
determined by high-resolution optical absorption
spectroscopy\cite{BBCD91}.
Furthermore, a fractional quantum number of pseudorotational angular
momentum due to Berry's phase\cite{Ham87,ZKP90} has been observed
in the rotational energy spectrum of Na$_3$\cite{DGWW86,ZwGr87}, supporting
the pseudorotation ground state.

A number of theoretical calculations\cite{PKPK83,WGLB87,BPFK87,KBK90,SuKa90}
predicted that lithium clusters are
planar up to Li$_6$.  However, the optical absorption spectrum for Li$_6$
appears to be consistent with the theoretical spectra
for a three-dimensional structure,
eliminating the possibility of planar geometries\cite{KFK91,DBKB91}.
Detailed geometrical parameters are still unknown.
For Li$_7$, EPR spectra\cite{GaLi84} were
interpreted as a pentagonal bipyramid ($D_{5h}$) in agreement with
theoretical prediction\cite{WGLB87,BPFK87,SuKa90,FKP84}.
As far as we know, no attempt to
identify the structure of Li$_5$ has been made.

Recently, Howard {\it et al.}\cite{HJJE93}
investigated Li$_5$ in an adamantane matrix
using EPR. The data indicate that all five nuclei are magnetically
identical and this is interpreted as a pseudorotating cluster.
Furthermore, the authors
suggested that the ground state of Li$_5$ is not a planar
but a three-dimensional structure, contrary to theoretical predictions.
A trigonal bipyramid, which is nearly as stable as the planar
structure\cite{BPFK87,KSW94},
is a prime candidate for the most stable
three-dimensional structure.
Since the pseudorotation
mechanism for a trigonal bipyramid molecule such as PF$_5$ is
known\cite{Berr60},  their speculation seems reasonable.
However, no conclusion can be made from the observed isotropic
spin population (ISP).
In order to extract
the structural information out of the experimental data,
quantitative theoretical calculations are desired.

In this paper, we report a pseudopotential local density functional
calculation of Li$_5$.  Two low-energy isomers, a trigonal
bipyramid ($^2B_1$)
and a planar structure ($^2A_1$), are studied as candidates for
the ground state.
The isotropic spin population is computed in order to identify
the structure observed by EPR.  Furthermore, the
possibility of two pseudorotation mechanisms is investigated by
searching the lowest potential barriers for permutation of the nuclei.

The computational method used in this work is the same as the
one previously used for
other systems. The details are given in Ref. \cite{KaWe91}.
The electron density is calculated using the local density
approximation (LDA)\cite{KoSh65}
with a generalized norm-conserving pseudopotential\cite{Hama89}. In order to
increase transferability, a fractionally occupied electronic configuration,
$1s^2 2s^{0.8} 2p^{0.2}$, is used to construct a
pseudopotential\cite{Hama89}.
A core radius $R_c=2.1$ is sufficiently accurate for
all orbitals\cite{KSW94}.
Wave functions are expanded in a plane wave basis set using a
simple cubic supercell.
Since Li$_3$ has a relatively large dipole moment, a large lattice
constant, $a$ = 60 a.u., was used to avoid a long-range dipole
interaction between repeated cells.  For other sizes, $a$ = 30 a.u.
was sufficiently large.
The same cutoff energy $E_{cut}$ = 11.2 Ry was used for all sizes.
This requires more than 130000 plane waves for the largest cell.
Both electronic and geometric structures are optimized by the steepest
descent method.  The ISP at the $i$th nuclear
position, $\vec{R}_i$ is evaluated as
\begin{equation}
\rho_i = \sum_\sigma \sum_n \frac{|\psi_{n,\sigma} (\vec{R}_i)|^2}
{|\psi_{2s}(0)|^2} \sigma
\end{equation}
where $\psi_{n,\sigma}$ is the $n$-th molecular orbital with a spin $\sigma
= \pm 1$
and $\psi_{2s}$
is the $2s$ orbital of an isolated Li atom.
All calculations have been carried out using the
CM-200 and CM-5 computers.

We have calculated Li$_3$, Li$_5$ and Li$_7$ clusters
at the same level of
accuracy.  The ISP's for Li$_3$
and Li$_7$ is compared
with the experimental data in order to provide the degree of
accuracy of our calculation for Li$_5$.

For Li$_3$, two equilibrium triangle structures, $^2B_2$ (an obtuse
isosceles triangle) and $^2A_1$ (an acute isosceles triangle) are
nearly degenerate as predicted in other calculations\cite{GeSc78,MCB83}.
The $^2B_2$ state appeared to be more
stable than the $^2A_1$ state.
However, the difference in total energy is only 8.3 meV.  The experimental
value, 3.2 meV \cite{BBCD91}, is within the accuracy
of the present calculation.
The conical intersection of two Born-Oppenheimer (BO) surfaces at
the $D_{3h}$ symmetric point
is only 66 meV (the experimental value = 33 meV) above the $^2B_2$ minimum,
indicating that the system is
quantum mechanically delocalized and forms a $D_{3h}$
instead of a $C_{2v}$ structure.
These results are consistent with
the pseudorotating state identified by the EPR
spectroscopy\cite{GaLi83,HSM84,Hish92}.
The ISP's for both extremum geometries are listed in Table \ref{TBL:li3a}.
The total ISP, $\rho_{total}$ is 0.48 and 0.78 for
the $^2A_1$ and $^2B_2$ states, respectively.   Since the
pseudorotating state occupies both extremum states,
the actual $\rho_{total}$
should be between these values.  Assuming equal weights on the two states,
$\rho_{total}$=0.63 is obtained, in good agreement with the
experimental value, $\rho_{total}$=0.69.

The EPR experiment determines the number of magnetically equivalent
nuclei and the isotropic spin population at each nucleus but it is
not capable of determining the detailed geometry such as equilibrium
bond distances.
Recently, the extremum geometries were precisely determined using
high-resolution photoabsorption spectroscopy through the analysis of
vibronic spectra\cite{BBCD91}.
The present calculation gives
excellent agreement with the experiment, as shown in Table \ref{TBL:li3b}.

Beyond the trimers, both optical and EPR spectra are so complicated
that the assignment or interpretation of spectra is almost impossible
without theoretical assistance.  For Li$_7$,
however,
two distinct groups of nuclei, two nuclei with large ISP
($\rho_1$=0.25) and five nuclei with very small
ISP ($\rho_2$=$-$0.015) were found from EPR\cite{GaLi84}, consistent
with the theoretically predicted pentagonal
bipyramid\cite{WGLB87,BPFK87,SuKa90,FKP84}.
The present calculation
predicts that two apical nuclei have $\rho_1$=0.23 and
five nuclei forming a pentagonal ring carry $\rho_2$=$-$0.01,
in good agreement with the experiment.

The results for Li$_3$ and Li$_7$ suggest that these calculations
are sufficiently accurate to predict the ISP
of small lithium clusters.  Therefore, the geometry of small clusters
can be determined by comparing theoretical ISP's with experimental data.

\begin{figure}
\psfig{figure=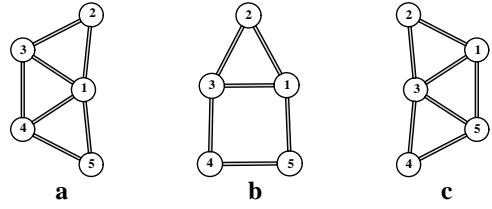,width=3in,angle=270}
\caption{Pseudorotation path for the planar Li$_5$ ($^2A_1$).
The center structure
corresponds to the transition state between two extremum states (the left and
the right figures).}
\label{FIG:2D}
\end{figure}

\begin{figure}
\psfig{figure=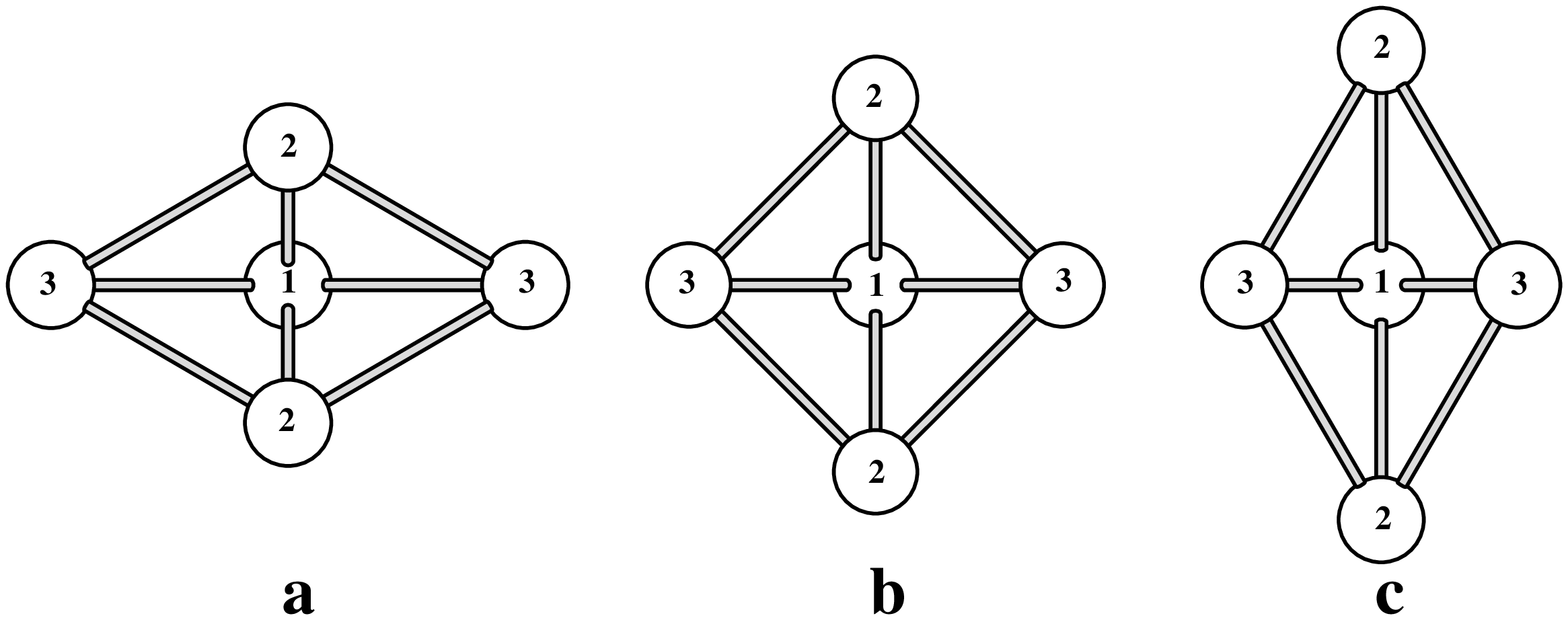,width=3in,angle=270}
\caption{Pseudorotation path for the trigonal bipyramid ($^2B_1$).
The center structure
corresponds to the transition state between two extremum states (the left and
the right figures) }
\label{FIG:3D}
\end{figure}

For Li$_5$,
most theoretical calculations\cite{PKPK83,WGLB87,BPFK87,KBK90,SuKa90}
predicted that the $^2A_1$ state of a
$C_{2v}$ planar structure [See Fig.\ \ref{FIG:2D}(a)] is the ground state.
However,
in contradiction to the previous calculations, we found that a distorted
trigonal bipyramid with the $C_{2v}$ symmetry
[See Fig. \ref{FIG:3D}(a)] is lower in total energy than
the $^2A_1$ state within the LDA.
This result seems consistent with the speculation by
Howard {\it et al.}\cite{HJJE93}.
However, the energy difference is only 46 meV/atom.
Since the accuracy in LDA total energy is not
always satisfactory\cite{RSOS93},
determining the ground state by the total energy alone is not reliable
if there are nearly degenerate states.

In Table \ref{TBL:li5a}, the theoretical ISP values of two structures are
compared with the experimental value.
The total ISP, $\rho_{total}=0.53$, for the trigonal bipyramid is
much smaller than the experimental value, 0.71.
On the other
hand, $\rho_{total} = 0.72$ for
the planar structure nearly coincides with
the experimental value.  This implies that the observed Li$_5$ cluster is
the planar cluster.
The small ISP for the trigonal
bipyramid is due to the large population of the $p$--character electron
which stabilizes this geometry significantly.  Large $sp$ mixing
is particular to Li clusters and does not appear in
other alkali-metal atom clusters.

The experiment also showed that the five nuclei are
magnetically equivalent.
Only a pentagonal ring ($D_{5h}$) has five equivalent nuclei
without dynamical transformation.  However, it is an unstable
state with total energy of 1.06 eV above the the
lowest $^2B_1$ state and, therefore, energetically not accessible.
The observed spectra must be due to the rapid transformation of
one of the low energy
structures.
In this paper, we assume the system
is pseudorotating by quantum-mechanical tunneling.

By analogy from the Berry pseudorotation of PF$_5$, the trigonal bipyramid
permutes its nuclei through the $C_{4v}$ transition state, as illustrated
in Fig.\ \ref{FIG:3D}.   For Li$_5$, however, since the unpaired
electron has a node
on the atoms 3-1-3 in Fig.\ \ref{FIG:3D}(a) and on atoms 2-1-2 in
Fig.\ \ref{FIG:3D}(c), the electronic wave functions before and after
the transition are orthogonal.  Therefore, tunneling
through the $C_{4v}$ point is essentially prohibited.
Furthermore, the potential barrier
height, 0.33 eV is too high to allow a rapid tunneling at low temperature.
The pseudorotation of the trigonal bipyramid is not likely to
occur within the experimental time scale.

Similarly to the pseudorotation of Li$_3$,
we expect that the planar
Li$_5$ pseudorotates around the $D_{5h}$ symmetric point if the potential
barrier to the cyclic permutation of nuclei is sufficiently low.
The $D_{5h}$ point is 0.82 eV above
the $^2A_1$ state, which
is too high to pass through even if the zero-point energy is taken into
account.
However, we found a relatively low transition state
to reach a nearest equivalent extremum in the planar structure.
The path from one extremum to
another is illustrated in Fig.\ \ref{FIG:2D}.
The barrier height,
0.18 eV, is higher than that of Li$_3$
but significantly lower than that of the trigonal bipyramid
pseudorotation path.
In contrast to the trigonal bipyramid, the nodal planes before and after
tunneling are nearly parallel and thus the tunneling is
allowed.
Although a nearly free pseudorotation like
the Li$_3$ case is not possible for Li$_5$, the tunneling probability
of the planar structure is expected to be much larger than that of
the trigonal bipyramid, supporting the idea that the planar structure
is the ground
state and in consistent with the CI calculation
(Table\ \ref{TBL:li5b}).

In order to make all nuclei equivalent, tunneling
has to be faster than the
observation time of the experiment.  Unfortunately,
a quantum Monte Carlo simulation
is necessary to determine the quantum-mechanical behavior of Li nuclei.
A recent quantum Monte Carlo simulation\cite{BaMi92} indicates that
zero-point energy and
tunneling play a crucial role even for large clusters.
Furthermore, the estimated zero-point energy for the bulk Li crystal is as
big as 33 meV/atom\cite{DaCo86}.  Since lithium is the third lightest
element, the quantum-mechanical behavior of the nuclei is not surprising.

\begin{figure}
\centerline{ \psfig{figure=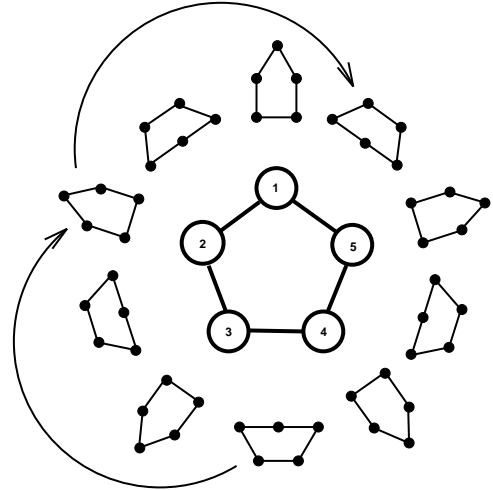,width=2.5in,angle=270} }
\vspace{0.2in}
\caption{Pseudorotation around the $D_{5h}$ symmetry point.  Arrows represent
the transition shown in Fig.\ \protect\ref{FIG:2D}.}
\label{FIG:circle}
\end{figure}

The dimension of the pseudorotation is 6 for Li$_5$ in contrast to 2 for
Li$_3$.
Because of this high dimensionality, the Li$_5$ pseudorotation is
not straightforward.
The permutation path described in Fig.\ \ref{FIG:2D} is actually
not a simple
rotation.  In terms of rotational motion, the transition is
not between the
nearest
neighbors as shown in Fig.\ \ref{FIG:circle} due to the high dimensionality
of the
system.  The path is a closed loop in the six-dimensional space and
its projection onto a two-dimensional plane rotates
three times (6$\pi$) around
the $D_{5h}$ symmetry point.
It is
interesting to calculate Berry's phase associated with the geometrical
transformation along the path\cite{ZKP90,Mead92}
and the fractional quantization of rotational
motion.  However, calculation of Berry's phase is not simple for
Li$_5$ because of its high dimensionality and the degeneracy at the $D_{5h}$
point\cite{ZKP90,Mead92}.

In conclusion, the theoretical isotropic spin population indicates that
the planar $C_{2v} (^2A_1)$ structure is more
consistent with the EPR
data than the $C_{2v} (^2B_1)$ trigonal bipyramid.  Furthermore, the
pseudorotation of the trigonal bipyramid is expected to be very slow
due to the symmetry of the electronic wave function and the high
potential barrier.
A new pseudorotation path
thorough a transition state at $C_{2v} (^2B_1)$ is proposed for the
planar structure.
The low potential barrier height probably permits rapid
tunneling between extrema in the BO surface.

\acknowledgements
We would like to acknowledge helpful discussions with T. Hamilton and
K. Lammertsma.
This work was partly supported by the Office of
Naval Research through Grant No. ONRN0014-91J-1835.
Calculations were performed on CM-5 and CM-200
at the Naval Research Laboratory.

\begin{table}
\caption{Isotropic spin populations for Li$_3$.  The experimental data
are for the pseudorotating state.}

\begin{tabular}{l|ccc|ccc}

 &\multicolumn{3}{c|}{$^2B_2$}&\multicolumn{3}{c}{$^2A_1$}\\
Method&$\rho_1$&$\rho_2$&$\rho_{total}$&$\rho_1$&$\rho_2$&$\rho_{total}$\\
\hline
CEPA$^a$&        0.02 & 0.35 & 0.72 & 0.22 & 0.19 & 0.60 \\
CI$^b$&          0.05 & 0.28 & 0.60 & 0.31 & 0.14 & 0.60 \\
PW-PP-LDA$^c$&   0.11 & 0.34 & 0.78 & 0.40 & 0.04 & 0.48 \\
Experiment$^d$&   --  &  --  & (0.69) &  --  &  --  & (0.69)

\end{tabular}

$^a$ Reference \onlinecite{GeSc78} \hspace{0.2in}
$^b$ Reference \onlinecite{Kend78} \hspace{0.2in}
$^c$ This work \\
$^d$ References \onlinecite{GaLi83,HSM84,Hish92}

\label{TBL:li3a}

\end{table}

\begin{table}
\caption{Comparison of calculated
extremal geometries of Li$_3$. The cohesive energy $E_{coh}$ is in eV,
relative energy to the $^2B_2$ state $\Delta E$ in eV,
side length $R$ in a.u., and apex angle $\theta$ in degrees.}

\begin{tabular}{l|ccc|ccc}

 &\multicolumn{3}{c|}{$^2B_2$}&\multicolumn{3}{c}{$^2A_1$}\\

Method&$E_{coh}$&$R$&$\theta$&$\Delta E$&$R$&$\theta$ \\
\hline
CEPA$^a$        & 0.50 & 5.23 & 72$^\circ$ &  0.010 & 5.69 & 54$^\circ$ \\
PP-LSD$^b$      & 0.49 & 5.3  & 73$^\circ$ &  0.00  & 5.1  & 52$^\circ$ \\
PW-PP-LDA$^c$   & 0.58 & 5.08 & 72$^\circ$ &  0.008 & 5.59 & 52$^\circ$ \\
Experiment$^d$  & 0.60 & 5.16 & 72$^\circ$ &  0.003 & 5.77 & 50$^\circ$ \\
\end{tabular}

$^a$ Reference \onlinecite{GeSc78} \hspace{0.2in}
$^b$ Reference \onlinecite{MCB83} \hspace{0.2in}
$^c$ Present work.\\
$^d$ References \onlinecite{BBCD91} and \onlinecite{Wu76}

\label{TBL:li3b}
\end{table}

\begin{table}
\caption{Isotropic spin populations $\rho_i$ for Li$_5$. See
Figs.\ \protect\ref{FIG:2D}(a) and \protect\ref{FIG:3D}(a) for
the location of the $i$th nucleus.}
\begin{tabular}{lcccc}

Structure&$\rho_1$&$\rho_2$&$\rho_3$&$\rho_{total}$\\
\hline
$^2A_1$&0.05&0.17&0.17&0.72\\
$^2B_1$&0.03&0.27&-0.03&0.53\\
Experiment&--&--&--&0.71

\end{tabular}
\label{TBL:li5a}
\end{table}

\begin{table}
\caption{Comparison of theoretical calculations for the
low-energy geometries
and cohesive energy of $Li_5$.  $E_{coh}$ is in eV, and
equilibrium bond distances in a.u.
The atom numbers are defined in Figs. \protect\ref{FIG:2D}(a)
and \protect\ref{FIG:3D}(a). for the $^2A_1$ and $^2B_1$ states,
respectively.}

\begin{tabular}{l|cccccc}
Method & Structure &$E_{coh}$&$R_{3-4}$&$R_{1-3}$&$R_{2-3}$&$R_{1-2}$ \\
\hline
HF$^a$             & $^2A_1$    & 0.51 & 5.49 & 6.65 & 5.50 & 5.49 \\
CI$^a$             & $^2A_1$    & 0.74 & 5.44 & 5.54 & 5.56 & 5.54 \\
CI$^b$             & $^2A_1$    & 0.60 & 5.82 & 5.82 & 5.82 & 5.84 \\
PW-PP-LDA$^c$      & $^2A_1$    & 0.80 & 5.65 & 5.21 & 5.47 & 5.44 \\
\hline
\hline
Method & Structure &$E_{coh}$&$R_{1-2}$&$R_{2-2}$&$R_{1-3}$&$R_{2-3}$ \\
\hline
CI$^b$             & $^2B_1$    & 0.56 & 5.31 & 6.41 & 5.90 & 6.20 \\
PW-PP-LDA$^c$      & $^2B_1$    & 0.85 & 4.81 & 5.43 & 5.38 & 5.58 \\
\end{tabular}
\label{TBL:li5b}

$^a$ Reference \onlinecite{RaJe85}\hspace{0.15in}
$^b$ Reference \onlinecite{BPFK87}\hspace{0.15in}
$^c$ This work
\end{table}

\begin{thebibliography}{10}

\bibitem{KFK91}
V. Bona{\u{c}}i{\'{c}}-Kouteck{\'{y}}, P. Fantucci, and J. Kouteck{\'{y}},
  Chem. Rev. {\bf 91},  1035  (1991).

\bibitem{BCDW90}
M. Broyer {\it et~al.}, Phys. Rev. A {\bf 42},  6954  (1990).

\bibitem{DBKB91}
P. Dugourd {\it et~al.}, Phys. Rev. Lett. {\bf 67},  2638  (1991).

\bibitem{BKBC92}
J. Blanc {\it et~al.}, J. Chem. Phys. {\bf 96},  1793  (1992).

\bibitem{GaLi83}
D.~A. Garland and D.~M. Lindsay, J. Chem. Phys. {\bf 78},  2813  (1983).

\bibitem{HSM84}
J.~A. Howard, R. Sutcliffe, and B. Mile, Chem. Phys. Lett. {\bf 112},  84
  (1984).

\bibitem{Hish92}
N. Hishinuma, Phys. Rev. A {\bf 46},  7023  (1992).

\bibitem{GeSc78}
W.~H. Gerber and E. Schumacher, J. Chem. Phys. {\bf 69},  1962  (1978).

\bibitem{MCB83}
J.~L. Martins, R. Car, and J. Buttet, J. Chem. Phys. {\bf 78},  5646  (1983).

\bibitem{BBCD91}
J. Blanc {\it et~al.}, Z. Phys. D {\bf 19},  7  (1991).

\bibitem{Ham87}
F.~S. Ham, Phys. Rev. Lett. {\bf 58},  725  (1987).

\bibitem{ZKP90}
J.~W. Zwanziger, M. Koenig, and A. Pines, Annu. Rev. Phys. Chem. {\bf 41},  601
   (1990).

\bibitem{DGWW86}
G. Delacr{\'{e}}taz {\it et~al.}, Phys. Rev. Lett. {\bf 56},  2598  (1986).

\bibitem{ZwGr87}
J.~W. Zwanziger and E.~R. Grant, J. Chem. Phys. {\bf 87},  2954  (1987).

\bibitem{PKPK83}
D. Plav{\u{s}}i{\'{c}}, J. Kouteck{\'{y}}, G. Pacchioni, and V.
  Bona{\u{c}}i{\'{c}}-Kouteck{\'{y}}, J. Phys. Chem. {\bf 87},  1096  (1983).

\bibitem{WGLB87}
Y. Wang, T.~F. George, D.~M. Lindsay, and A.~C. Beri, J. Chem. Phys. {\bf 86},
  3493  (1987).

\bibitem{BPFK87}
I. Boustani {\it et~al.}, Phys. Rev. B {\bf 35},  9437  (1987).

\bibitem{KBK90}
J. Kouteck{\'{y}}, I. Boustani, and V. Bona{\u{c}}i{\'{c}}-Kouteck{\'{y}}, Int.
  J. Quantum Chem. {\bf 38},  149  (1990).

\bibitem{SuKa90}
O. Sugino and H. Kamimura, Phys. Rev. Lett. {\bf 65},  2696  (1990).

\bibitem{GaLi84}
D.~A. Garland and D.~M. Lindsay, J. Chem. Phys. {\bf 80},  4761  (1984).

\bibitem{FKP84}
P. Fantucci, J. Kouteck{\'{y}}, and G. Pacchioni, J. Chem. Phys. {\bf 80},  325
   (1984).

\bibitem{HJJE93}
J.~A. Howard {\it et~al.}, Chem. Phys. Lett. {\bf 204},  128  (1993).

\bibitem{KSW94}
R. Kawai, M.-W. Sung, and J.~H. Weare (unpublished).

\bibitem{Berr60}
R.~S. Berry, Rev. Mod. Phys. {\bf 32},  447  (1960).

\bibitem{KaWe91}
R. Kawai and J.~H. Weare, J. Chem. Phys. {\bf 95},  1151  (1991).

\bibitem{KoSh65}
W. Kohn and L. Sham, Phys. Rev. {\bf A140},  1133  (1965).

\bibitem{Hama89}
D.~R. Hamann, Phys. Rev. B {\bf 40},  2980  (1989).

\bibitem{RSOS93}
K. Raghavachari {\it et~al.}, Chem. Phy. Lett. {\bf 214},  357  (1993).

\bibitem{BaMi92}
P. Ballone and P. Milani, Phys. Rev. B {\bf 45},  11222  (1992).

\bibitem{DaCo86}
M.~M. Dacorogna and M.~L. Cohen, Phys. Rev. B {\bf 34},  4996  (1986).

\bibitem{Mead92}
C.~A. Mead, Rev. Mod. Phys. {\bf 64},  51  (1992).

\bibitem{Kend78}
J. Kendrick, Mol. Phys. {\bf 35},  593  (1978).

\bibitem{Wu76}
C.~H. Wu, J. Chem. Phys. {\bf 65},  3181  (1976).

\bibitem{RaJe85}
B.~K. Rao and P. Jena, Phys. Rev. B {\bf 32},  2058  (1985).

\end{thebibliography}
\end{document}